\documentclass[aps,prc,preprint,showpacs,showkeys]{revtex4}
\usepackage{amssymb}
\usepackage{amsmath}
\usepackage{graphicx}
\usepackage{bm}
\usepackage{epsf}

\begin{document}

\title{Impact of the symmetry energy on nuclear pasta phases and crust-core
transition in neutron stars}
\author{S. S. Bao}
\affiliation{School of Physics, Nankai University, Tianjin 300071, China}
\author{H. Shen}~\email{shennankai@gmail.com}
\affiliation{School of Physics, Nankai University, Tianjin 300071, China}

\begin{abstract}
We study the impact of the symmetry energy on properties of nuclear pasta
phases and crust-core transition in neutron stars. We perform a
self-consistent Thomas--Fermi calculation employing the relativistic
mean-field model. The properties of pasta phases presented in the inner
crust of neutron stars are investigated and the crust-core transition
is examined. It is found that the slope of the symmetry
energy plays an important role in determining the pasta phase structure
and the crust-core transition. The correlation between the symmetry energy
slope and the crust-core transition density obtained in the Thomas--Fermi
approximation is consistent with that predicted by the liquid-drop model.
\end{abstract}

\maketitle


\section{Introduction}
\label{sec:1}

Neutron stars are great laboratories for the study of asymmetric nuclear
matter over a wide range of density. With increasing depth in the
neutron star, the matter density may rise from the subnuclear region
to several times normal nuclear matter density~\cite{Lattimer04,Chamel08}.
In general, a neutron star consists of an outer crust of nuclei in a
gas of electrons, an inner crust of nuclei in a gas of neutrons and
electrons, and a liquid core of uniform dense matter~\cite{Chamel08,PR00,Webe05}.
The inner crust of neutron stars has drawn much attention due to its
complex phase structure and significant role in astrophysical
observations~\cite{Stei08,Rave83,Gril12,Okmaoto13}.
In the inner crust, spherical nuclei may become unstable
as the density increases toward the crust-core transition, and the stable
nuclear shape is likely to change from droplet to rod, slab, tube, and bubble,
known as nuclear pasta phases~\cite{Rave83,Gril12,Okmaoto13,Mene08,Bao14}.
The crust-core transition occurs at the density where the energy
density of the homogeneous matter becomes lower than that of the pasta phase.
During the last decades, the properties of pasta phases have been investigated by
using various methods, such as the liquid-drop model~\cite{Rave83,Wata00,Bao14}
and the Thomas--Fermi approximation~\cite{Gril12,Mene08,Oyam07,Mene10}.
In Ref.~\cite{Oyam07}, the density region of nonspherical nuclei was evaluated by
using a parametrized Thomas--Fermi approximation, which was found to be sensitive
to the density dependence of the nuclear symmetry energy. In Ref.~\cite{Gril12},
a self-consistent Thomas--Fermi approximation was used to calculate properties
of the inner crust including pasta phases, and it was found that the symmetry
energy and its slope could have significant impacts on the pasta phase
structure and crust-core transition. In our previous work~\cite{Bao14},
the effects of the symmetry energy on pasta phase properties and crust-core
transition were investigated by employing the coexisting phases method
based on a liquid-drop model, and the correlation between the symmetry
energy slope and the crust-core transition was obtained and analyzed.
It is noticeable that the symmetry energy and its slope could
play an important role in determining the pasta phase structure and
crust-core transition in neutron stars.

In recent years, the nuclear symmetry energy and its density dependence
have received great interest due to their importance for understanding
many phenomena in nuclear physics and astrophysics~\cite{PR07,LiBA08,Horo01,Duco10}.
The value of the symmetry energy $E_{\rm sym}$ at saturation density
is constrained by experiments to be about $30\pm 4$ MeV,
while its slope $L$ at saturation density
is still very uncertain and may vary from about $20$
to $115$ MeV~\cite{Chen13}. It has been found that various properties
of neutron stars, such as the crust structure, the crust-core transition,
and the star radius, are sensitive to the symmetry energy $E_{\rm sym}$
and its slope $L$~\cite{Gril12,Oyam07,Duco11,Mene11}.
In Ref.~\cite{Duco11}, the impact of the symmetry energy on the crust-core
transition was examined using various effective Skyrme and relativistic
approaches, in which the crust-core transition density obtained from
the dynamical and thermodynamical methods showed a clear decrease with
increasing $L$. The correlation between the crust-core transition
density and the symmetry energy slope $L$ has been extensively studied
in the literature using various methods~\cite{Gril12,Oyam07,Bao14,Duco11}.
It is shown that the resulting transition density depends on the method
and effective nuclear interaction used in the calculation.
It is important to make further investigations in order to determine
a clear correlation between the symmetry energy slope $L$ and the
crust-core transition.

The main purpose of this article is to investigate the impact of the
symmetry energy on pasta phase properties and explore the correlation
between the symmetry energy slope $L$ and the crust-core transition.
We perform a self-consistent Thomas--Fermi calculation employing the
relativistic mean-field (RMF) model~\cite{Sero86,Meng06} for nuclear
interactions. In the Thomas--Fermi approximation,
the surface effect and nucleon distributions are treated self-consistently,
rather than by assuming a sharp interface as in the coexisting phases
method~\cite{Bao14}. In our most recent study~\cite{Shen14}, we made
a detailed comparison between the Thomas--Fermi (TF) approximation and
the coexisting phases (CP) method with only the droplet configuration.
It is interesting to compare their difference in pasta phases including
all configurations as mentioned above.
For the nuclear interaction, we adopt the RMF model with two different
parametrizations, TM1~\cite{TM1} and IUFSU~\cite{IUFSU}, both of
which are known to be successful in describing the ground-state properties
of finite nuclei, including unstable ones.
In the RMF approach, nucleons interact via the exchange of scalar and
vector mesons, and the model parameters are generally fitted to nuclear
matter saturation properties or ground-state properties of finite nuclei.
The TM1 parametrization includes nonlinear terms for both $\sigma$ and
$\omega$ mesons, while an additional $\omega$-$\rho$ coupling term
is added in the IUFSU parametrization. It is well known that
the $\omega$-$\rho$ coupling term plays a crucial role in
modifying the density dependence of the symmetry energy and affecting
the neutron star properties~\cite{Mene11,IUFSU,Horo01,Horo03,Prov13}.
In order to evaluate the impact of the symmetry energy slope $L$ on
pasta phase properties and crust-core transition, we employ two sets of
generated models based on the TM1 and IUFSU parametrizations as given in
Ref.~\cite{Shen14}. The model parameters were determined by simultaneously
adjusting $g_{\rho}$ and ${\Lambda}_{\rm{v}}$ so as to achieve a
given $L$ at saturation density $n_0$ while keeping $E_{\rm{sym}}$ fixed at
a density of 0.11 fm$^{-3}$. We note that the fixed density in Ref.~\cite{Bao14}
was chosen to be the saturation density, namely, $n_{\rm{fix}}=n_0$,
whereas $n_{\rm{fix}}$ = 0.11 fm$^{-3}$ was used in Ref.~\cite{Shen14}.
It has been shown in Ref.~\cite{Shen14} that the choice of the fixed density
$n_{\rm{fix}}$ = 0.11 fm$^{-3}$ could produce very similar binding
energies for finite nuclei within one set of generated models.
Furthermore, all models in each set have the same isoscalar saturation
properties and fixed symmetry energy at $n_{\rm{fix}}$ = 0.11 fm$^{-3}$,
but they have different symmetry energy slope $L$.
By using the set of models with different $L$, it is possible to study
the impact of $L$ on pasta phase properties and explore the correlation
between $L$ and the crust-core transition.

This article is organized as follows. In Sec.~\ref{sec:2},
we briefly describe the RMF model and the self-consistent
TF approximation used in this study.
In Sec.~\ref{sec:3}, we present the numerical results and examine
the impact of the symmetry energy on pasta phase properties,
while the correlation between the symmetry energy slope $L$ and
the crust-core transition is discussed.
Section~\ref{sec:4} is devoted to the conclusions.

\section{ Formalism}
\label{sec:2}

The inner crust of neutron stars is studied within the TF
approximation by employing the RMF model for nuclear interactions.
In the RMF model~\cite{Sero86,Meng06}, nucleons interact through
the exchange of various mesons. The mesons considered here are
the isoscalar-scalar meson $\sigma$, the isoscalar-vector meson $\omega$,
and the isovector-vector meson $\rho$.
For a system consisting of protons, neutrons, and electrons,
the Lagrangian density reads
\begin{eqnarray}
\label{eq:LRMF}
\mathcal{L}_{\rm{RMF}} & = & \sum_{i=p,n}\bar{\psi}_i
\left\{i\gamma_{\mu}\partial^{\mu}-\left(M+g_{\sigma}\sigma\right)
-\gamma_{\mu} \left[g_{\omega}\omega^{\mu} +\frac{g_{\rho}}{2}\tau_a\rho^{a\mu}
+\frac{e}{2}\left(1+\tau_3\right)A^{\mu}\right]\right\}\psi_i  \notag \\
& & +\bar{\psi}_{e}\left[i\gamma_{\mu}\partial^{\mu} -m_{e} +e \gamma_{\mu}
A^{\mu} \right]\psi_{e}  \notag \\
&& +\frac{1}{2}\partial_{\mu}\sigma\partial^{\mu}\sigma -\frac{1}{2}%
m^2_{\sigma}\sigma^2-\frac{1}{3}g_{2}\sigma^{3} -\frac{1}{4}g_{3}\sigma^{4}
\notag \\
&& -\frac{1}{4}W_{\mu\nu}W^{\mu\nu} +\frac{1}{2}m^2_{\omega}\omega_{\mu}%
\omega^{\mu} +\frac{1}{4}c_{3}\left(\omega_{\mu}\omega^{\mu}\right)^2  \notag
\\
&& -\frac{1}{4}R^a_{\mu\nu}R^{a\mu\nu} +\frac{1}{2}m^2_{\rho}\rho^a_{\mu}%
\rho^{a\mu} +\Lambda_{\rm{v}} \left(g_{\omega}^2
\omega_{\mu}\omega^{\mu}\right)
\left(g_{\rho}^2\rho^a_{\mu}\rho^{a\mu}\right) -\frac{1}{4}%
F_{\mu\nu}F^{\mu\nu},
\end{eqnarray}
where $W^{\mu\nu}$, $R^{a\mu\nu}$, and $F^{\mu\nu}$ are the antisymmetric field
tensors corresponding to $\omega^{\mu}$, $\rho^{a\mu}$, and $A^{\mu}$, respectively.
In the RMF approach, the meson fields are treated as classical fields,
and the field operators are replaced by their expectation values.
For a static system, the nonvanishing expectation values are
$\sigma =\left\langle \sigma \right\rangle$,
$\omega =\left\langle\omega^{0}\right\rangle$,
$\rho =\left\langle \rho^{30} \right\rangle$,
and $A =\left\langle A^{0}\right\rangle$.
The equations of motion for these mean fields derived from the Lagrangian
density~(\ref{eq:LRMF}) have the following form:
\begin{eqnarray}
&&-\nabla ^{2}\sigma +m_{\sigma }^{2}\sigma +g_{2}\sigma ^{2}+g_{3}\sigma
^{3}=-g_{\sigma }\left( n_{p}^{s}+n_{n}^{s}\right) ,
\label{eq:eqms} \\
&&-\nabla ^{2}\omega +m_{\omega }^{2}\omega +c_{3}\omega^{3}
+2\Lambda_{\rm{v}}g^2_{\omega}g^2_{\rho}{\rho}^2 \omega
=g_{\omega}\left( n_{p}+n_{n}\right) ,
\label{eq:eqmw} \\
&&-\nabla ^{2}\rho +m_{\rho }^{2}{\rho}
+2\Lambda_{\rm{v}}g^2_{\omega}g^2_{\rho}{\omega}^2{\rho}
=\frac{g_{\rho }}{2}\left(n_{p}-n_{n}\right) ,
\label{eq:eqmr} \\
&&-\nabla ^{2}A=e\left( n_{p}-n_{e}\right) ,
\label{eq:eqma}
\end{eqnarray}%
where $n_i^s$ and $n_i$ denote, respectively, the scalar and
number densities of species $i$.
The equations of motion for nucleons give the standard relations
between the densities and chemical potentials,
\begin{eqnarray}
\mu _{p} &=&{\sqrt{\left( k_{F}^{p}\right)^{2}+{M^{\ast }}^{2}}}+g_{\omega
}\omega +\frac{g_{\rho }}{2}\rho +e A,
\label{eq:mup} \\
\mu _{n} &=&{\sqrt{\left( k_{F}^{n}\right)^{2}+{M^{\ast }}^{2}}}+g_{\omega
}\omega -\frac{g_{\rho }}{2}\rho ,
\label{eq:mun}
\end{eqnarray}%
where $M^{\ast}=M+g_{\sigma}\sigma$ is the effective nucleon mass,
and $k_{F}^{i}$ is the Fermi momentum of species $i$, which is related
to the number density by $n_i=\left(k_{F}^{i}\right)^3/3\pi^2$.

The matter in the inner crust of neutron stars contains protons,
neutrons, and electrons under the conditions of $\beta$ equilibrium
and charge neutrality. We employ the Wigner--Seitz cell approximation
to describe the inner crust, in which the equilibrium state is
determined by minimization of the total energy density at zero temperature.
The stable cell shape may change from droplet to rod, slab, tube,
and bubble as the density increases. For simplicity, we assume the
electron density is uniform throughout the Wigner--Seitz cell,
since the electron screening effect is known to be negligible at
subnuclear densities~\cite{Maru05}. Furthermore, we also neglect the correction
caused by the Coulomb interaction with charged particles in other cells,
which is negligibly small in most cases~\cite{Oyam93,Shen11}.
In the TF approximation, the total energy per cell is calculated from
\begin{equation}
E_{\rm{cell}}=\int_{\rm{cell}}{\varepsilon}_{\rm{rmf}}({\bf r})d {\bf r}
  +{\varepsilon}_e V_{\rm{cell}},
\label{eq:TFe}
\end{equation}%
where ${\varepsilon}_e$ is the electron kinetic energy density, and
${\varepsilon}_{\rm{rmf}}({\bf r})$ is the local energy density at
position ${\bf r}$, which is given in the RMF model by
\begin{eqnarray}
{\varepsilon }_{\rm{rmf}}({\bf r}) &=&\displaystyle{\sum_{i=p,n}\frac{1}{\pi ^{2}}%
\int_{0}^{k_{F}^{i}}dk\,k^{2}\,\sqrt{k^{2}+{M^{\ast }}^{2}}}  \notag \\
&&+\frac{1}{2}(\nabla \sigma )^{2}+\frac{1}{2}m_{\sigma }^{2}\sigma ^{2}+%
\frac{1}{3}g_{2}\sigma ^{3}+\frac{1}{4}g_{3}\sigma ^{4}  \notag \\
&&-\frac{1}{2}(\nabla \omega )^{2}-\frac{1}{2}m_{\omega }^{2}\omega ^{2}-%
\frac{1}{4}c_{3}\omega ^{4}+g_{\omega }\omega \left( n_{p}+n_{n}\right)
\notag \\
&&-\frac{1}{2}(\nabla \rho )^{2}-\frac{1}{2}m_{\rho }^{2}\rho ^{2}
-\Lambda_{\rm{v}}g_{\omega }^{2}g_{\rho }^{2}\omega ^{2}\rho ^{2}+\frac{g_{\rho }}{2}%
\rho \left( n_{p}-n_{n}\right)   \notag  \\
&&-\frac{1}{2}(\nabla A)^{2}+eA\left( n_{p}-n_{e}\right).
\label{eq:ETF}
\end{eqnarray}%
Here, we consider different pasta configurations, including
the droplet, rod, slab, tube, and bubble.
The volume of the Wigner--Seitz cell for different configurations
can be written as
\begin{equation}
V_{\rm{cell}}=\left\{
\begin{array}{ll}
\frac{4}{3}{\pi}r_{\text{ws}}^{3} &\hspace{0.5cm} \textrm{(droplet and bubble)},  \\
l{\pi}r_{\text{ws}}^{2}           &\hspace{0.5cm} \textrm{(rod and tube)},        \\
2r_{\text{ws}}{l^2}               &\hspace{0.5cm} \textrm{(slab)},
\end{array}
\right.
\label{eq:vcell}
\end{equation}
where $r_{\text{ws}}$ is the radius of a spherical cell for the droplet
and bubble configurations, while the rod and tube have cylindrical shapes
with radius $r_{\text{ws}}$ and length $l$, and the slab has width $l$
and thickness $2r_{\text{ws}}$.
We note that the choices of the length for a cylindrical shape and the width
for a slab are somewhat arbitrary~\cite{Gril12}, which would not affect
the resulting energy density of the system.

At a given average baryon density $n_{b}$, we minimize the total
energy density with respect to the cell size $r_{\rm{ws}}$
for each pasta configuration, and then we compare the energy densities
between different configurations in order to
determine the most stable shape that has the lowest energy density.
Furthermore, the energy density of the corresponding homogeneous phase
at the same $n_{b}$ is also computed, and the crust-core transition
occurs at the density where the energy density of the homogeneous
phase becomes lower than that of the pasta phase.
In order to calculate the total energy per cell given by Eq.~(\ref{eq:TFe})
at fixed $r_{\rm{ws}}$ and $n_{b}$, we solve the coupled
Eqs.~(\ref{eq:eqms})--(\ref{eq:eqma}) under the constraints
of $\beta$ equilibrium, charge neutrality, and baryon number
conservation, which have the following form:
\begin{eqnarray}
\mu _{n} &=&\mu _{p}+\mu _{e},
\label{eq:beta} \\
N_{e} &=&N_{p}=\int_{\rm{cell}}n_{p}({\bf r})d {\bf r},
\label{eq:charge}\\
n_{b} V_{\rm{cell}}  &=& \int_{\rm{cell}}
  \left[ n_{p}({\bf r})+n_{n}({\bf r})\right] d {\bf r}.
\label{eq:nb}
\end{eqnarray}
In practice, we start with an initial guess for the mean fields
$\sigma ({\bf r})$, $\omega ({\bf r})$, $\rho ({\bf r})$, and $A({\bf r})$,
then determine the chemical potentials $\mu_{n}$, $\mu_{p}$,
and $\mu_{e}$ by the constraints~(\ref{eq:beta})--(\ref{eq:nb}).
Once the chemical potentials are obtained,
it is easy to calculate various densities and solve
Eqs.~(\ref{eq:eqms})--(\ref{eq:eqma}) to get new mean fields.
This procedure should be iterated until convergence is achieved.

\section{Results and discussion}
\label{sec:3}

\begin{figure}[htb]
\includegraphics[bb=17 14 541 758, width=7 cm,clip]{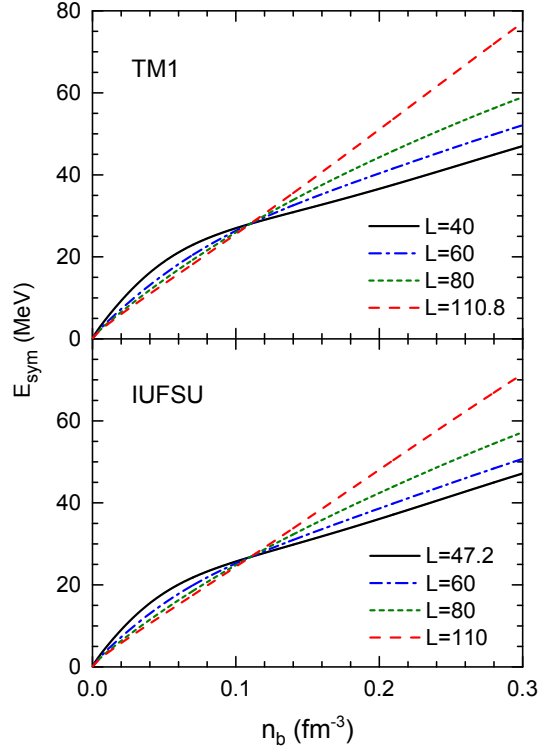}
\caption{(Color online) Symmetry energy $E_{\text{sym}}$ as a function of
the baryon density $n_b$ for modified versions of TM1 (upper panel)
and IUFSU (lower panel) with several values of $L$ at saturation density.
The symmetry energy is fixed at a density of 0.11 fm$^{-3}$.}
\label{fig:1esym}
\end{figure}

In this section, we present numerical results for the inner crust of
neutron stars, and we discuss the impact of the symmetry energy on
pasta phase properties and crust-core transition.
The results obtained from the self-consistent TF calculation are
compared with those obtained using the CP method~\cite{Bao14}.
For the effective nuclear interaction, we consider two successful
RMF models, TM1~\cite{TM1} and IUFSU~\cite{IUFSU}.
The parameter sets and saturation properties of these two models
are given in Tables~\ref{tab:1} and~\ref{tab:2}, respectively.
In order to clarify the correlation between the symmetry
energy slope $L$ and the crust-core transition, we employ two sets
of generated models based on the TM1 and IUFSU parametrizations as
given in Ref.~\cite{Shen14}. It is noticeable that all models in
each set have the same isoscalar saturation properties and fixed
symmetry energy $E_{\rm{sym}}$ at a density of 0.11 fm$^{-3}$ but have
different symmetry energy slope $L$. These models have been generated
by simultaneously adjusting $g_{\rho}$ and ${\Lambda}_{\rm{v}}$
so as to achieve a given $L$ at saturation density $n_0$ while
keeping $E_{\rm{sym}}$ fixed at a density of 0.11 fm$^{-3}$
as described in Ref.~\cite{Shen14}.
The parameters, $g_{\rho}$ and ${\Lambda}_{\rm{v}}$, generated from
the TM1 and IUFSU models for different $L$ are given in
Tables~\ref{tab:3} and~\ref{tab:4}, respectively.
In Fig.~\ref{fig:1esym}, we plot the symmetry energy $E_{\rm{sym}}$
as a function of the baryon density $n_b$ for the two sets of models
generated from TM1 (upper panel) and IUFSU (lower panel).
One can see that all models in each set have the same $E_{\text{sym}}$
at a density of 0.11 fm$^{-3}$, but they have different values of
$E_{\text{sym}}$ at lower and higher densities due to the difference in
the slope $L$. It is obvious that a smaller $L$ corresponds to
a larger (smaller) $E_{\text{sym}}$ at lower (higher) densities.
It will be shown below that the behavior of $E_{\text{sym}}$
plays a crucial role in determining the pasta phase structure
and the crust-core transition.

\begin{figure}[htb]
\includegraphics[bb=14 310 580 591, width=13 cm,clip]{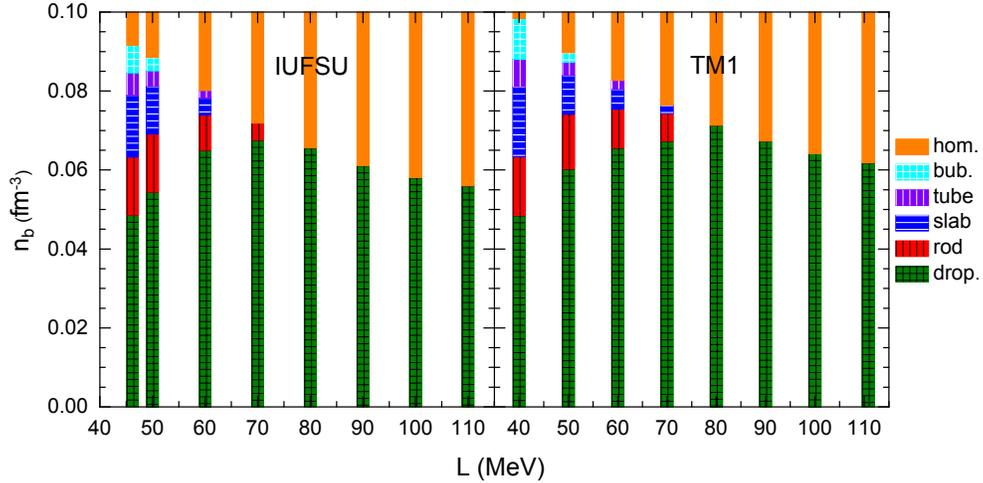}
\caption{(Color online) Phase diagrams for the two sets of models generated
from IUFSU (left panel) and TM1 (right panel). Different colors represent
droplet, rod, slab, tube, bubble, and homogeneous phases as indicated
in the legend.}
\label{fig:2diag}
\end{figure}

\begin{center}
\begin{figure}[thb]
\centering
\begin{tabular}{ccc}
\includegraphics[bb=27 181 572 677, width=0.32\linewidth, clip]{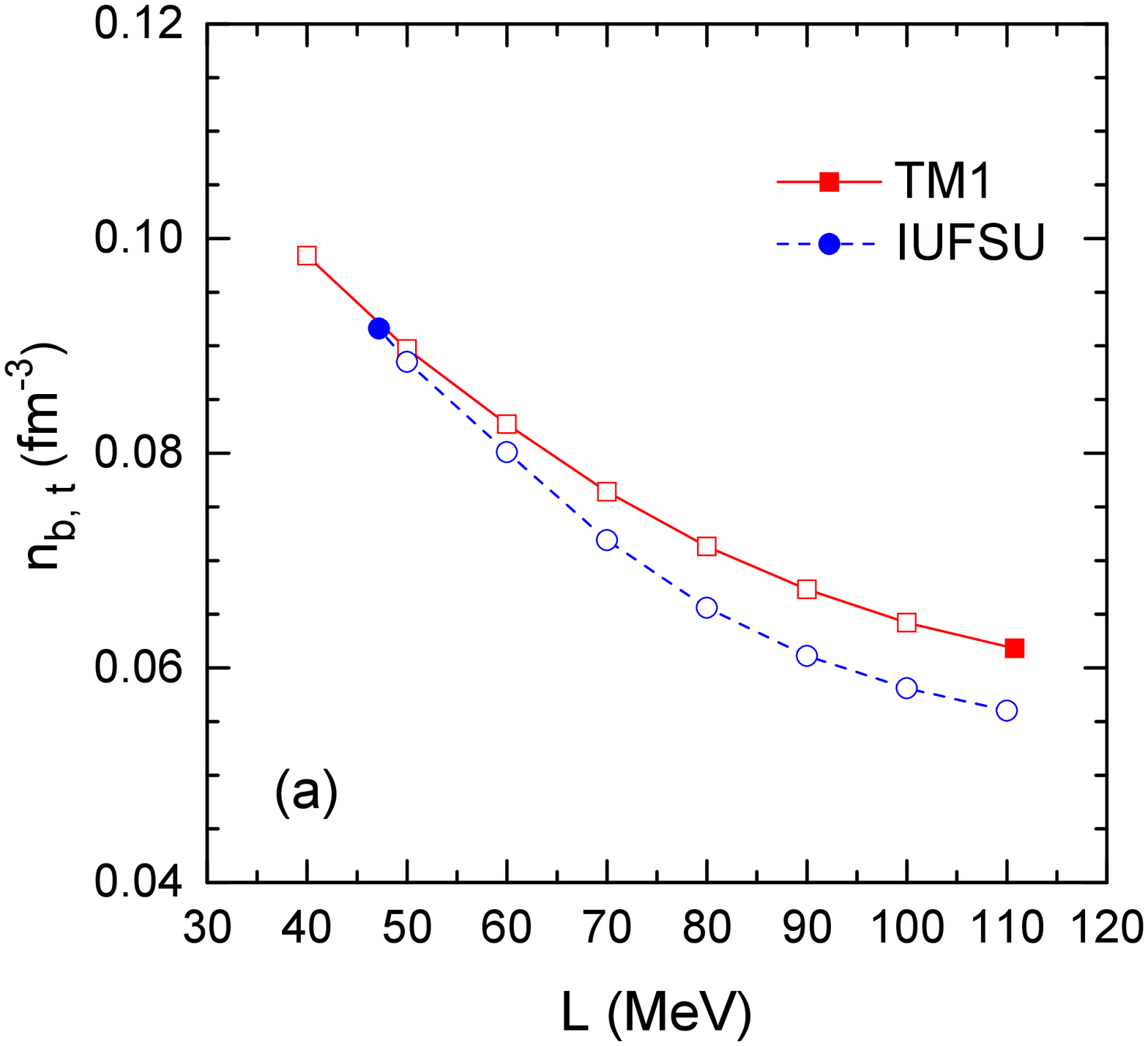}&
\includegraphics[bb=27 181 572 677, width=0.32\linewidth, clip]{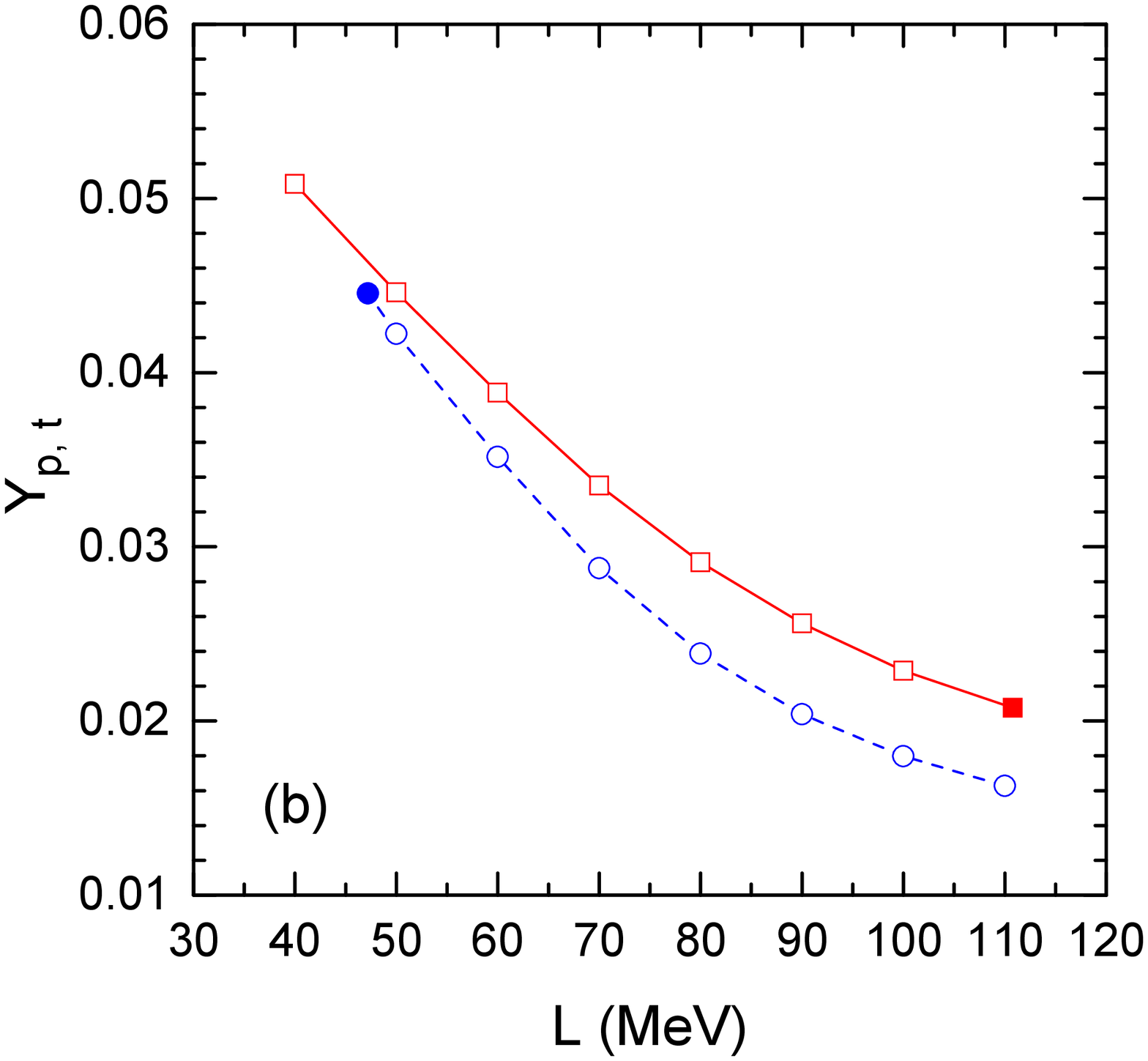}&
\includegraphics[bb=27 181 572 677, width=0.32\linewidth, clip]{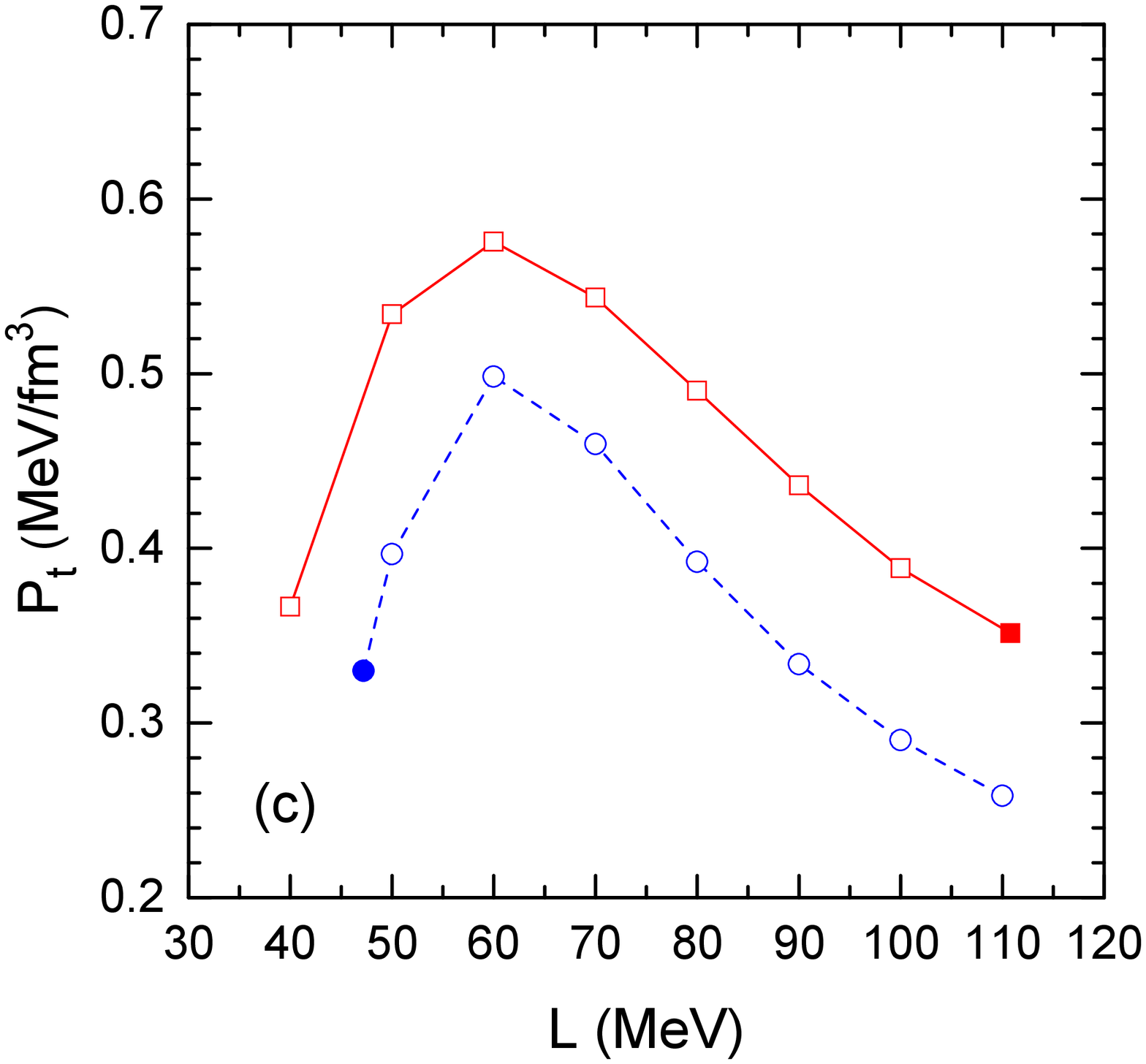}  \\
\end{tabular}
\caption{(Color online) Crust-core transition density $n_{b,t}$ (a),
proton fraction at the transition point, $Y_{p,t}$ (b), and pressure
at the transition point, $P_t$ (c), as a function of $L$ obtained in the TF
approximation using the two sets of models generated from TM1 (red-solid
line with squares) and IUFSU (blue-dashed line with circles).
The results obtained with the original TM1 and IUFSU models are indicated
by the filled square and circle, respectively.}
\label{fig:9PT}
\end{figure}
\end{center}

We first present the phase diagram for the inner crust of neutron stars
and discuss the influence of the symmetry energy slope $L$ on the pasta
phase structure. In Fig.~\ref{fig:2diag}, the density ranges of various
pasta phases obtained from the self-consistent TF calculation
are displayed for the two sets of generated models,
IUFSU (left panel) and TM1 (right panel).
It is found that only the droplet configuration can occur before the
crust-core transition for $L\geq 80$ MeV, whereas the pasta phase
structure may change from droplet to rod, slab, tube, and bubble
for smaller values of $L$ (e.g., $L=50$ MeV).
As one can see from Fig.~\ref{fig:2diag},
the onset density of nonspherical nuclei, i.e., the transition
density from droplet to rod, significantly decreases with decreasing $L$
in the low-$L$ region ($L\leq 70$ MeV).
This behavior is consistent with that reported in Ref.~\cite{Oyam07}.
It can be understood from a fission-like instability of spherical
nuclei predicted in the liquid-drop model~\cite{Oyam07,Iida02},
in which a spherical liquid drop becomes unstable to quadrupolar
deformations when the volume fraction of the liquid drop
reaches the fission-instability criterion, $u=(r_d/r_{\rm{ws}})^3=1/8$,
with $r_d$ and $r_{\rm{ws}}$ being, respectively, the radii of the droplet
and the Wigner--Seitz cell. From Fig.~\ref{fig:4rcd} below,
we can see that a smaller $L$ corresponds to a larger value of
$r_d/r_{\rm{ws}}$ in the droplet phase at low densities, and, therefore,
the model with a smaller $L$ results in an earlier onset of
nonspherical nuclei, as shown in Fig.~\ref{fig:2diag}.
On the other hand, the onset density of homogeneous matter
significantly decreases with increasing $L$, which has also been
observed in earlier studies~\cite{Gril12,Oyam07,Bao14,Duco10}.
This trend may be understood from the energy-density
curvature of pure neutron matter. According to the analysis
in the liquid-drop model~\cite{Duco10}, the energy-density
curvature of pure neutron matter at saturation density,
$C_{\rm{NM}}(n_0)$, is approximately proportional to $L$,
and the inhomogeneous phase occurs in the spinodal region of asymmetric
nuclear matter where the energy density has a negative curvature.
A rough analysis implies that the larger $C_{\rm{NM}}(n_0)$ is,
the farther away a spinodal border of $\beta$-equilibrium matter deviates
from the saturation density $n_0$. Therefore, the model with
a larger $L$ leads to a smaller onset density of homogeneous matter,
as shown in Fig.~\ref{fig:2diag} [see also Fig.~\ref{fig:9PT}(a)].
We conclude that a smaller value of $L$
results in an earlier onset of nonspherical nuclei and a later
transition to homogeneous matter. Hence, the model with a smaller
$L$ predicts a more complex phase structure and a larger density range of
pasta phases in neutron star crusts.

\begin{figure}[htb]
\includegraphics[bb=31 218 576 676, width=7 cm,clip]{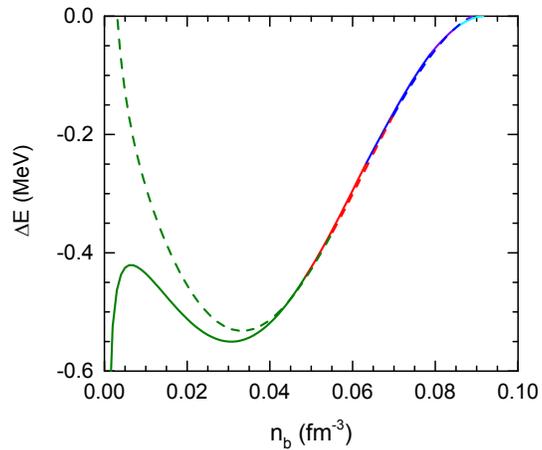}
\caption{(Color online) Energy per nucleon of the pasta phase relative to
that of homogeneous matter, $\Delta E$, as a function of $n_b$
obtained from the TF (solid lines) and CP (dashed lines) calculations
using the original IUFSU model. Different colors represent
different pasta phases as shown in Fig.~\ref{fig:2diag}.}
\label{fig:3ea}
\end{figure}

The phase diagram obtained in the present TF calculation
is very similar to that obtained using the CP method~\cite{Bao14}.
However, the bubble configuration could not appear even with
the lowest $L$ by using the CP method~\cite{Bao14}.
In order to make a detailed comparison of the pasta phase structure
between the TF and CP calculations, we present in Table~\ref{tab:5}
the onset densities of pasta phases for the original IUFSU ($L=47.2$ MeV)
and TM1 ($L=110.8$ MeV) models. It is shown that only the droplet
configuration appears in the case of TM1 for both TF and CP,
which is due to its very large value of $L$.
On the other hand, the original IUFSU model predicts that
almost all pasta phases can occur before the crust-core transition
due to its small value of $L$.
From Table~\ref{tab:5}, we can see that there are visible
differences in the onset densities between TF and CP.
This is mainly caused by the different treatments of surface and
Coulomb energies~\cite{Shen14}, which play an essential role in
determining the phase shape.
It is found that the onset density of homogeneous matter, i.e.,
the crust-core transition density, obtained in the TF approximation
is slightly higher than that obtained using the CP method.
This is because the configuration space of TF is much larger than
that of CP, and, therefore, a lower energy density can be achieved
in the minimization procedure of the TF calculation,
which leads to a higher crust-core transition density.
In Fig.~\ref{fig:3ea}, we plot the energy per nucleon of the
pasta phase relative to that of homogeneous matter, $\Delta E$,
as a function of the average baryon density, $n_b$, obtained from
the TF and CP calculations using the original IUFSU model.
One can see that there are significant differences in $\Delta E$
between TF and CP at lower densities, whereas the differences become
much smaller in the pasta phase region. The comparison between TF
and CP at low density with the droplet configuration has been
extensively discussed in our previous work~\cite{Shen14},
and it was found that the simple CP method could not describe
the nonuniform matter around the neutron drip density due to its
energy being higher than that of homogeneous matter.
From Fig.~\ref{fig:3ea}, it is seen that $\Delta E$ of the CP calculation
becomes positive at $n_b$ $<$ 0.003 fm$^{-3}$, which is consistent
with the onset density of the droplet phase given in Table~\ref{tab:5}.
On the other hand, the crust-core transition occurs at relatively
high density, beyond which $\Delta E$ becomes positive.
Although the differences in $\Delta E$ between TF and CP calculations
at higher densities are rather small (on the order of a few keV),
it may lead to visible differences in the onset densities
as shown in Table~\ref{tab:5}. We note that the differences in the
energy per nucleon between different pasta shapes are of the order
of 0.1--1 keV, which has also been reported in Refs.~\cite{Wata00,Oyam93}.
Therefore, the pasta phase structure is very sensitive
to the method and nuclear interaction used in the calculation.

In the present study, we focus on the correlation between the
symmetry energy slope $L$ and pasta phase properties.
In Fig.~\ref{fig:4rcd}, the size of the Wigner--Seitz cell,
$r_{\rm{ws}}$, and that of the inner part (nucleus or hole), $r_d$,
which are obtained from the TF calculation with two extreme values of $L$
in the TM1 (upper panel) and IUFSU (lower panel) sets,
are displayed as a function of $n_b$.
In the TF approximation, there is no distinct boundary between
the liquid phase and the gas phase, so we prefer to
define the size of the inner part, $r_d$, by density fluctuations,
similar to that of Refs.~\cite{Okmaoto13,Maru05}, as
\begin{equation}
r_{d}=\left\{
\begin{array}{ll}
r_{\rm{ws}} \left(\frac{\langle n_{p}\rangle^{2}}{\langle n_{p}^{2}\rangle}\right)^{1/D}
& \hspace{0.5cm} \textrm{(droplet, rod, and slab),} \\
r_{\rm{ws}}\left(1-\frac{\langle n_{p}\rangle ^{2}}{\langle n_{p}^{2}\rangle}\right)^{1/D}
& \hspace{0.5cm} \textrm{(tube and bubble),}%
\end{array}%
\right.
\end{equation}
with $D=1,2,3$ being the geometrical dimension of the system.
The average values, $\langle n_{p}\rangle$ and $\langle n_{p}^{2}\rangle$,
are calculated over the cell volume $V_{\rm{cell}}$ given by Eq.~(\ref{eq:vcell}).
From Fig.~\ref{fig:4rcd}, one can see that the model with the smallest $L$
yields a rather complex structure of the inner crust, whereas only the
droplet phase appears in the case with the largest $L$. It is found that
$r_{\rm{ws}}$ and $r_d$ show significant jumps at the transition points
between different pasta shapes, which indicates that the transition is
first order. One can see that $r_{\rm{ws}}$ decreases with $n_b$
at lower densities, while it rapidly increases before the crust-core
transition in the case of small $L$.
A similar behavior of $r_{\rm{ws}}$ was also observed in Ref.~\cite{Gril12}.

\begin{center}
\begin{figure}[thb]
\centering
\begin{tabular}{ccc}
\includegraphics[bb=33 172 574 676, width=0.32\linewidth, clip]{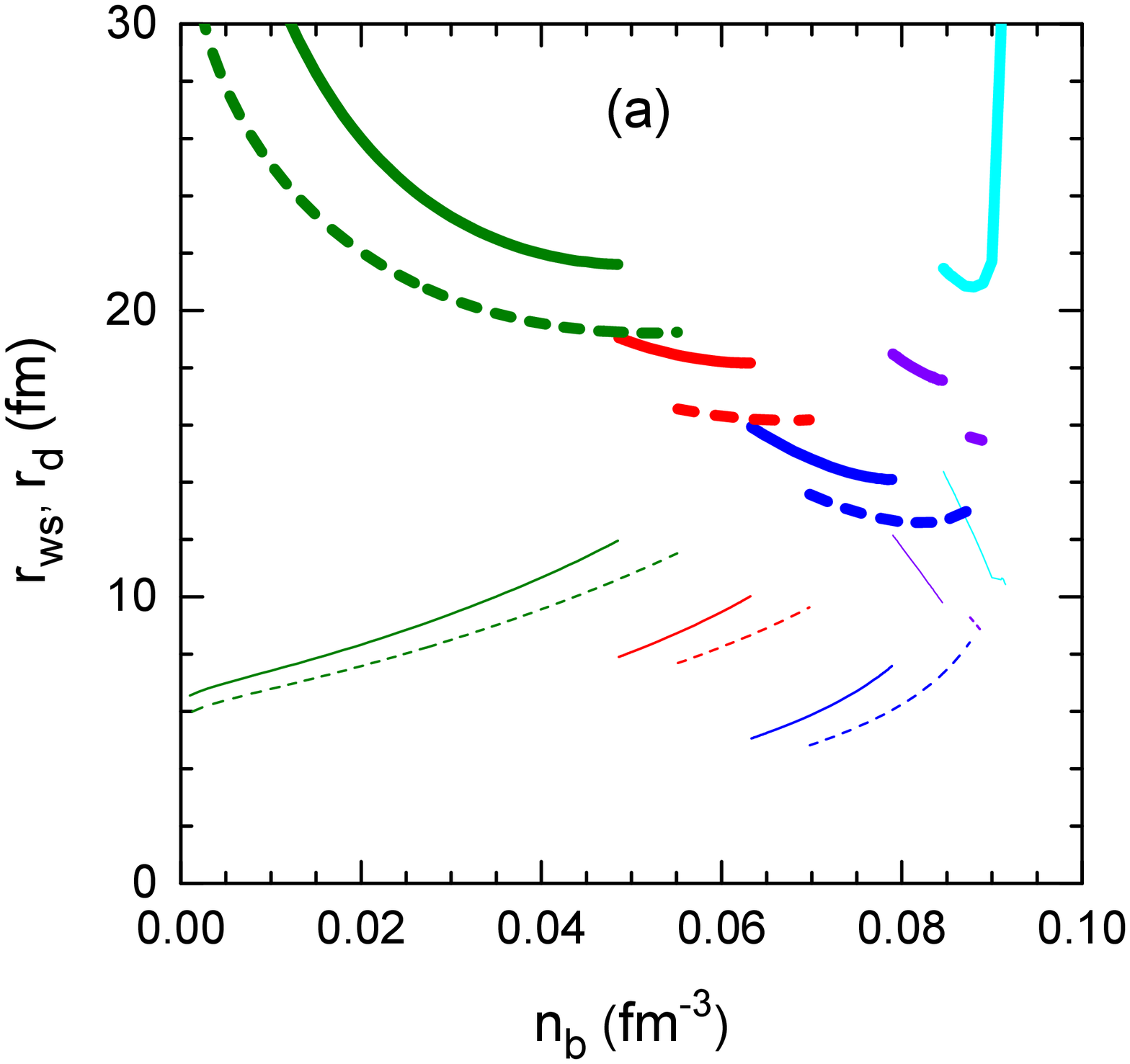}&
\includegraphics[bb=33 172 574 676, width=0.32\linewidth, clip]{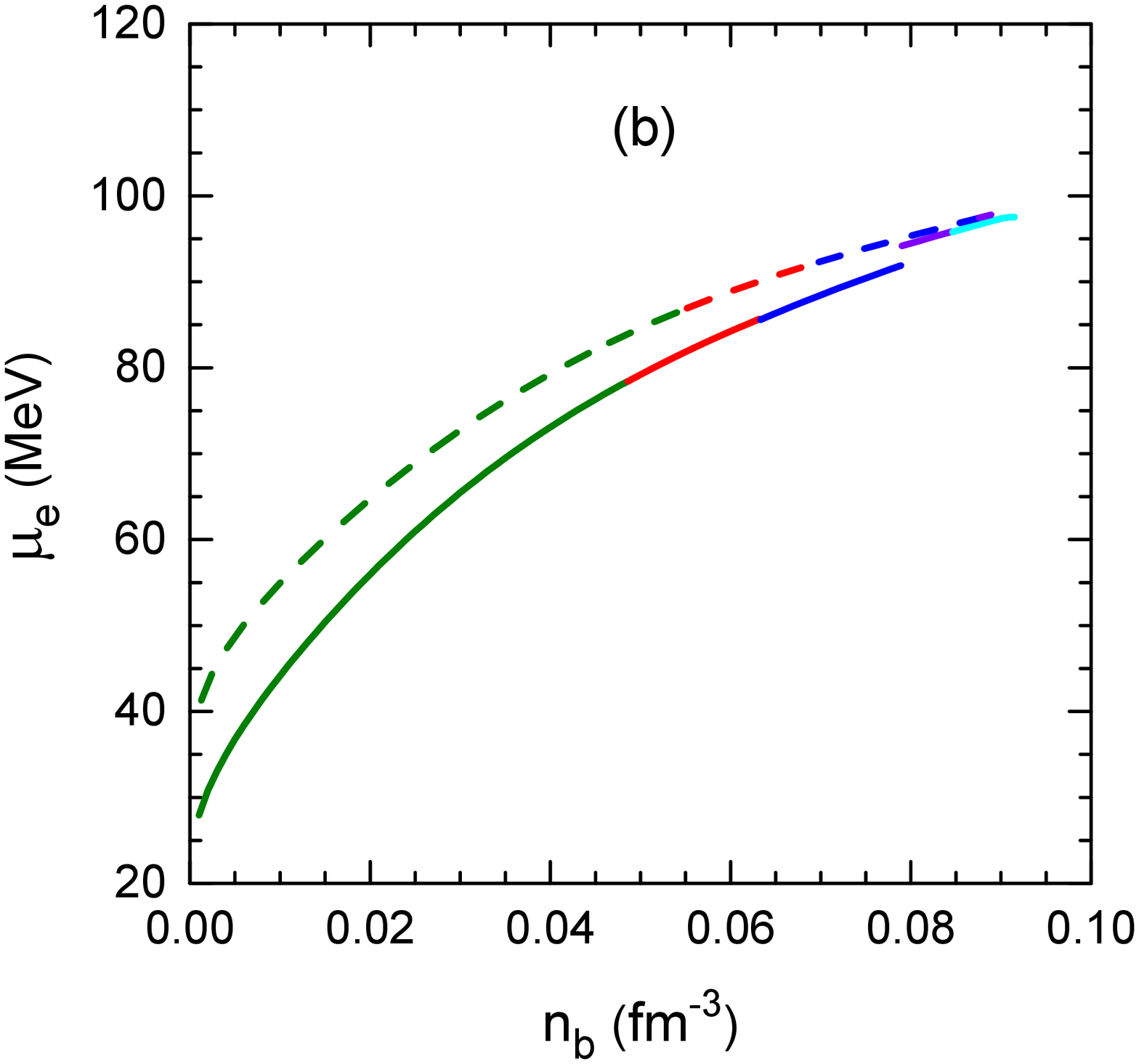}&
\includegraphics[bb=33 172 574 676, width=0.32\linewidth, clip]{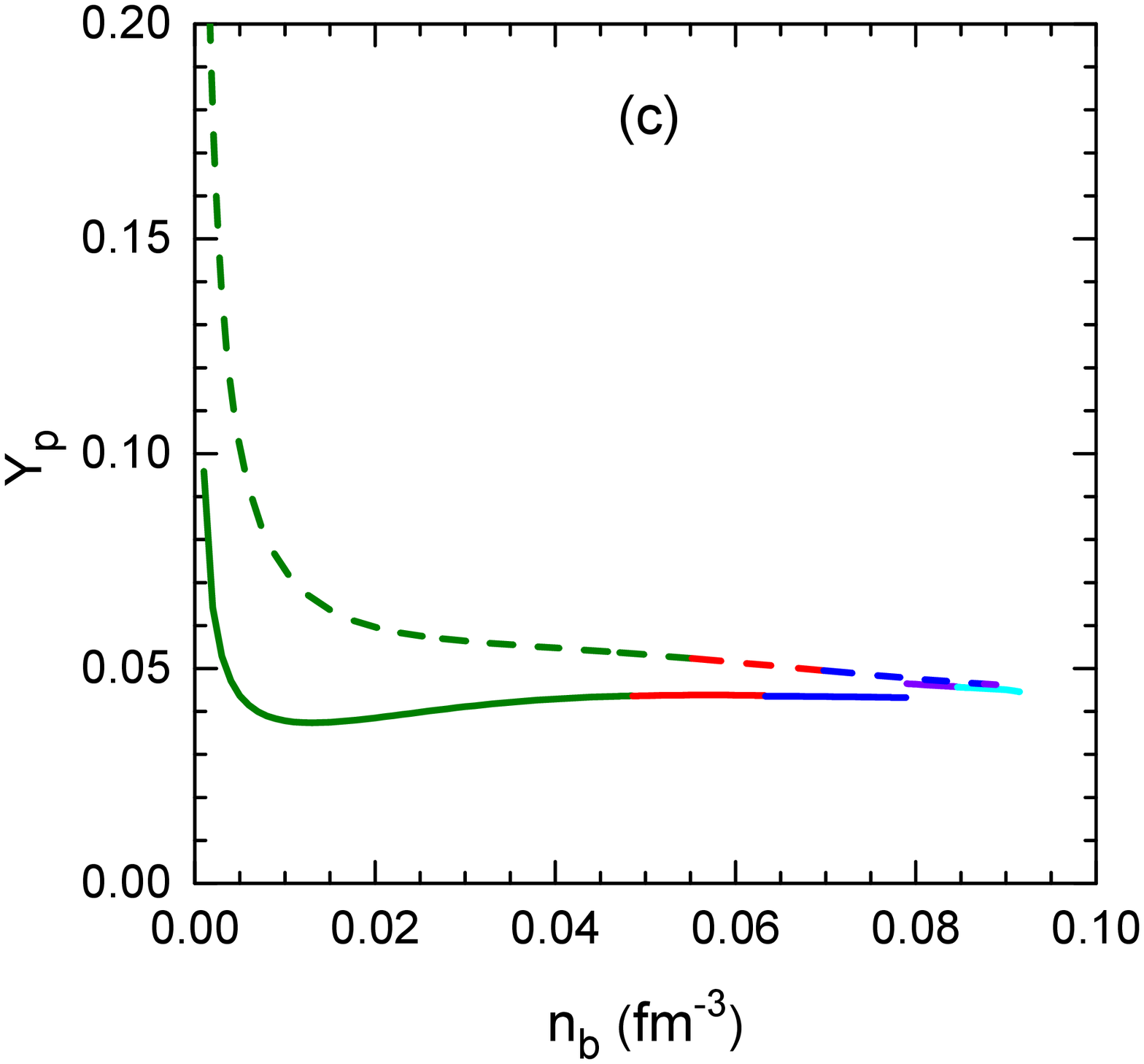}  \\
\end{tabular}
\caption{(Color online) Comparison of equilibrium properties of the inner crust
between the TF (solid lines) and CP (dashed lines) methods using the original
IUFSU model.
The size of the Wigner--Seitz cell, $r_{\rm{ws}}$ (thick lines), and that of the
inner part, $r_d$ (thin lines) (a), the electron chemical potential $\mu_{e}$ (b),
and the average proton fraction $Y_p$ (c) are plotted as a function of $n_b$.
The different colors represent different pasta phases.}
\label{fig:5CPTF}
\end{figure}
\end{center}

It is of interest to compare the results between the TF and CP methods.
In Fig.~\ref{fig:5CPTF}(a), $r_{\rm{ws}}$ and $r_d$ obtained in the TF
approximation are compared with those obtained by the CP method
using the original IUFSU model. It is found that both $r_{\rm{ws}}$ and
$r_d$ from the CP method are somewhat smaller than those from the TF method.
This can be explained by the behaviors of the electron chemical
potential $\mu_e$ shown in Fig.~\ref{fig:5CPTF}(b) and the average
proton fraction $Y_p$ shown in Fig.~\ref{fig:5CPTF}(c).
As discussed in Ref.~\cite{Shen14}, the CP method could yield
larger $\mu_e$ and $Y_p$ in comparison to those of TF
[see Figs.~\ref{fig:5CPTF}(b) and~\ref{fig:5CPTF}(c)].
This is because the surface and Coulomb energies are treated
perturbatively in the CP method, whereas they are included
self-consistently in the minimization of the TF method.
Due to larger $\mu_e$ and $Y_p$, the CP method would give rise
to smaller $r_{\rm{ws}}$ as shown in Fig.~\ref{fig:5CPTF}(a),
which could be explained by their correlation in the
liquid-drop model~\cite{Shen14}.
From Figs.~\ref{fig:5CPTF}(b) and~\ref{fig:5CPTF}(c),
one can see that there are small jumps
at $n_b \sim 0.079\, \rm{fm}^{-3}$ in the TF case,
which are caused by the transition from slab to tube.
It is obvious that trends in the differences between CP and TF
are very similar for all pasta phases, and they are
consistent with those obtained in the droplet phase~\cite{Shen14}.

\begin{center}
\begin{figure}[thb]
\centering
\begin{tabular}{ccc}
\includegraphics[bb=24 10 574 823, width=0.32\linewidth, clip]{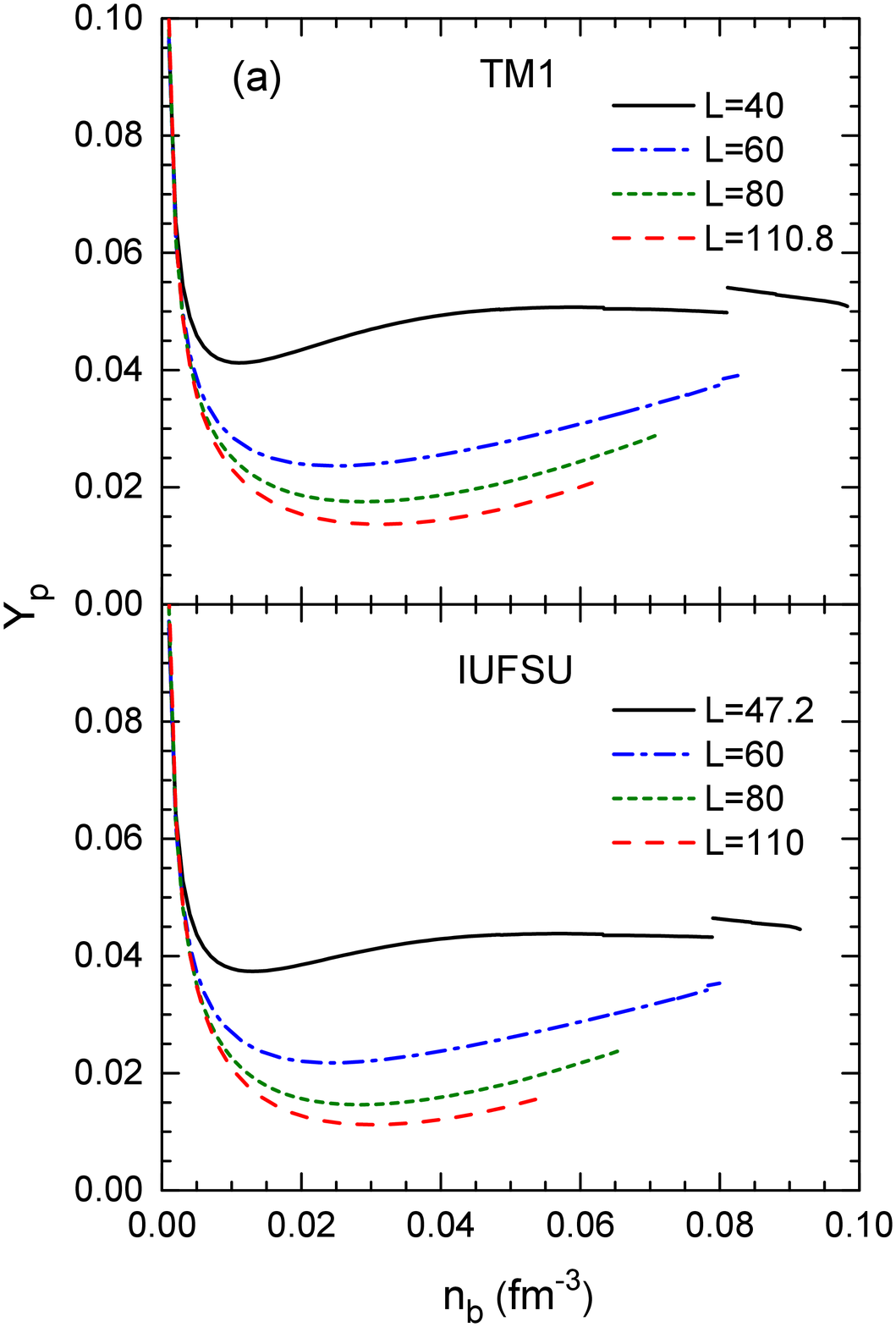}&
\includegraphics[bb=24 10 574 823, width=0.32\linewidth, clip]{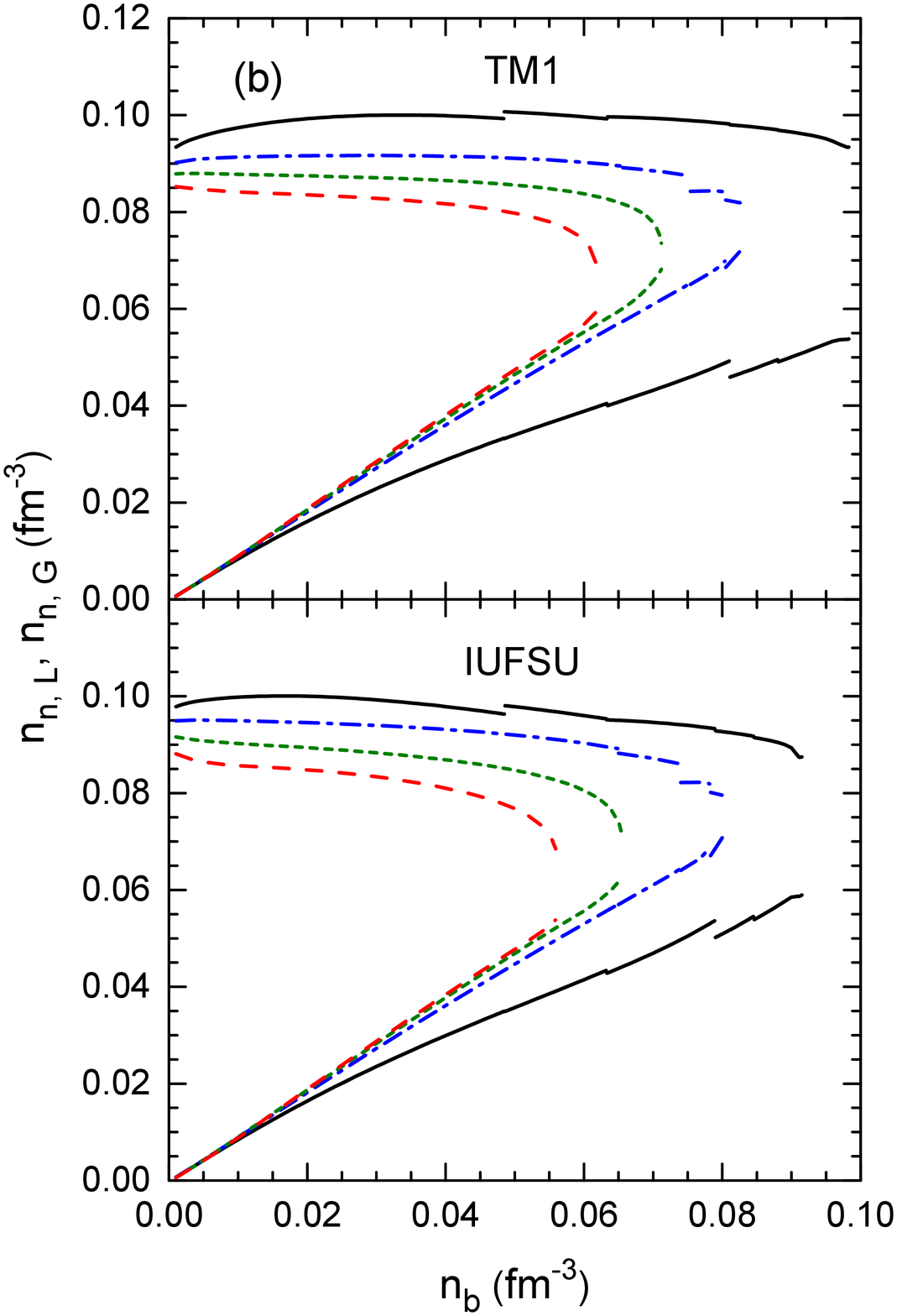}&
\includegraphics[bb=24 10 574 823, width=0.32\linewidth, clip]{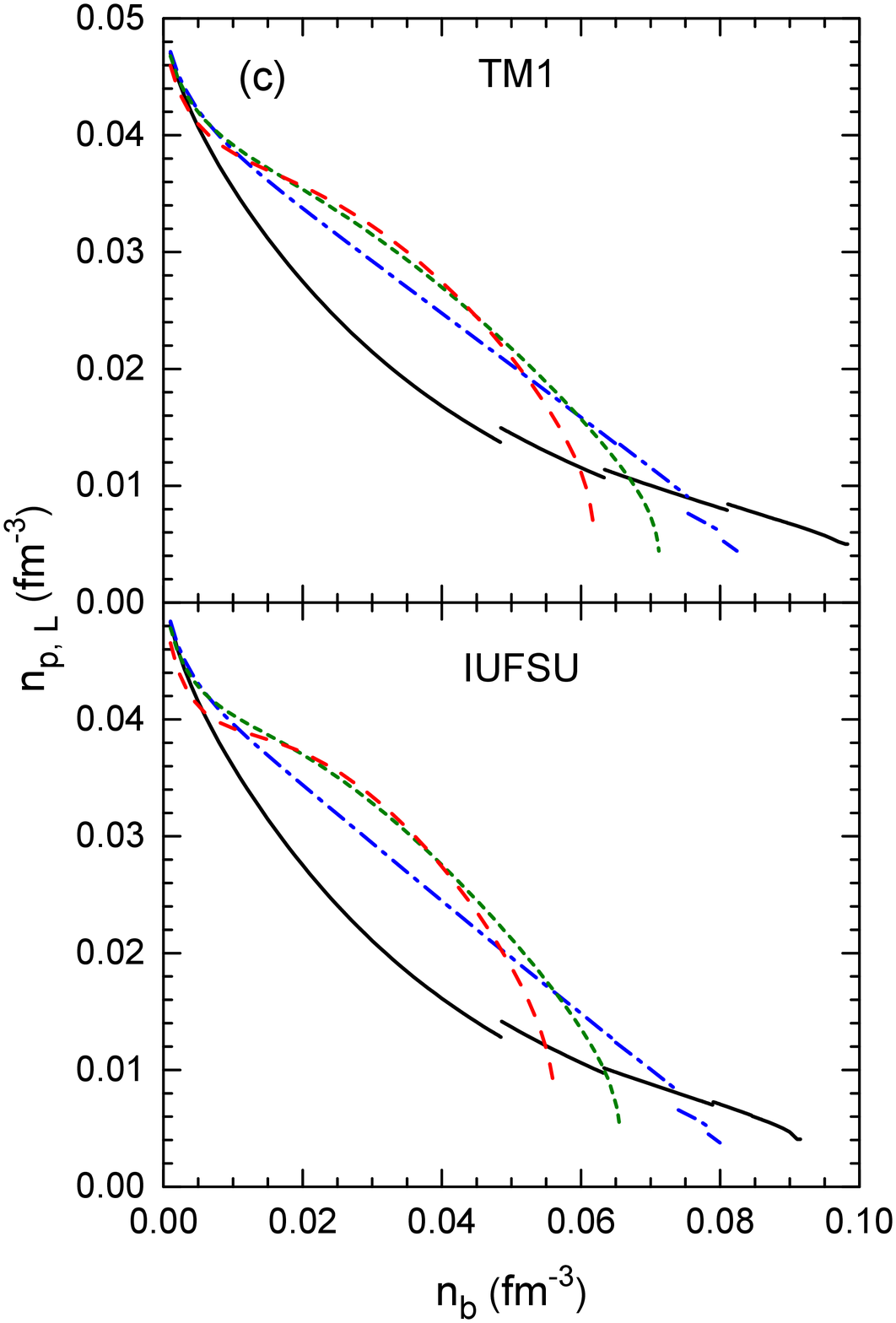}  \\
\end{tabular}
\caption{(Color online) Equilibrium properties of the Wigner--Seitz cell
obtained in the TF approximation for modified versions of TM1 (upper panel)
and IUFSU (lower panel) with several values of $L$.
The average proton fraction $Y_p$ (a), the neutron densities of the liquid phase
and the gas phase, $n_{n,L}$ and $n_{n,G}$, at the center or boundary of the cell (b),
and the proton density of the liquid phase, $n_{p,L}$, at the center or boundary
of the cell (c) are plotted as a function of $n_b$.
Small jumps are observed at the transition between different pasta shapes.}
\label{fig:6TF}
\end{figure}
\end{center}

\begin{center}
\begin{figure}[thb]
\centering
\begin{tabular}{ccc}
\includegraphics[bb=31 11 575 823, width=0.32\linewidth, clip]{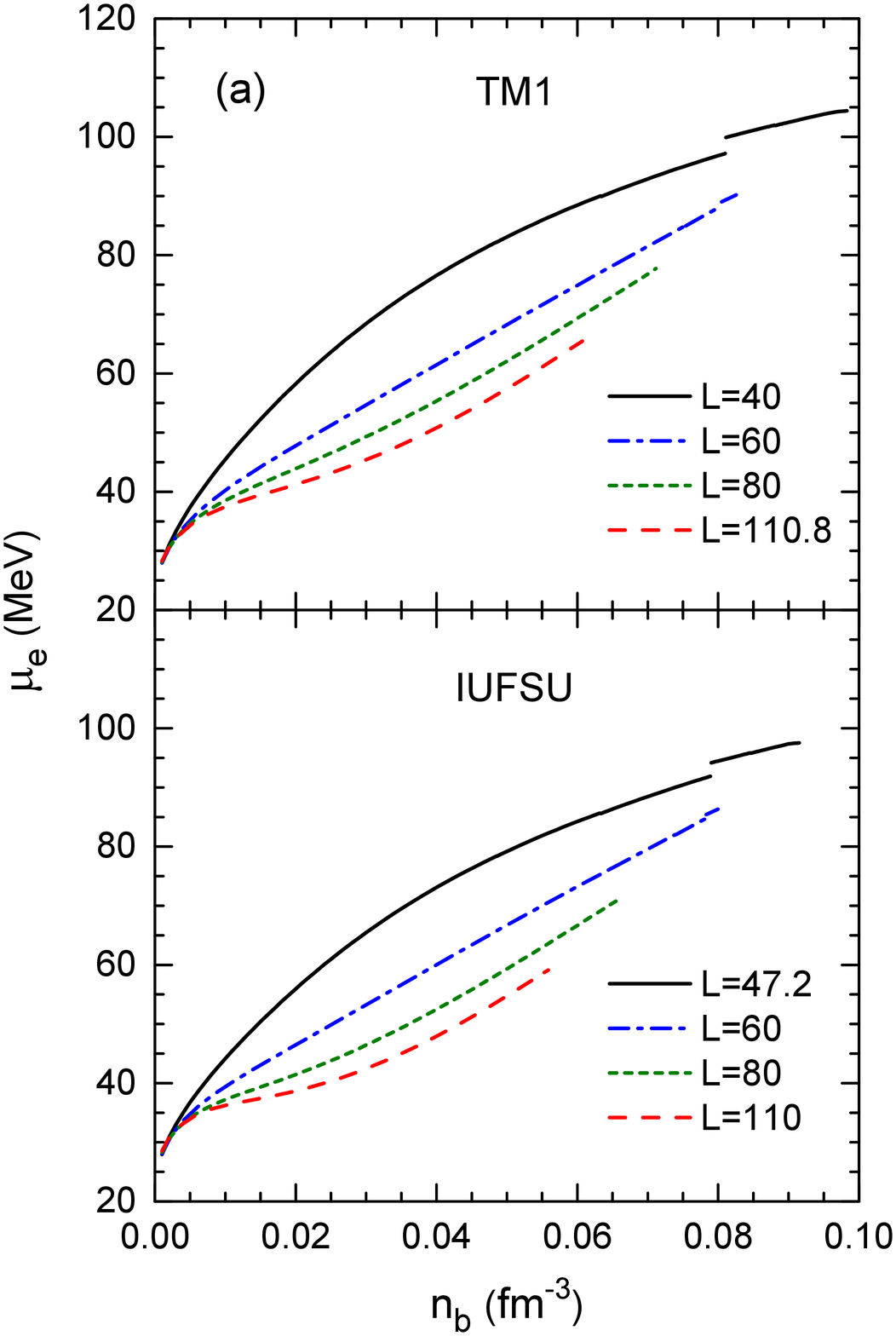}&
\includegraphics[bb=31 11 575 823, width=0.32\linewidth, clip]{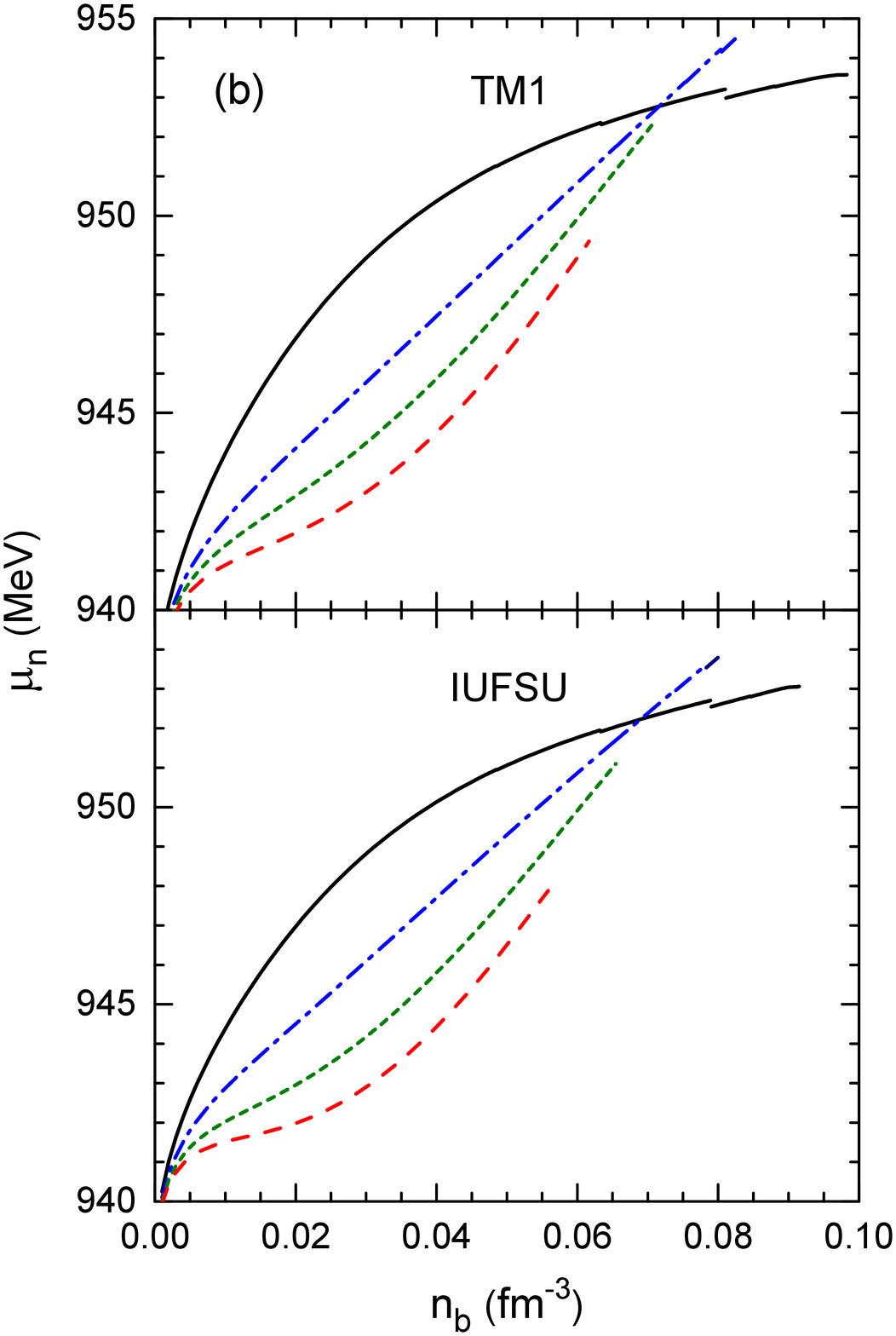}&
\includegraphics[bb=31 11 575 823, width=0.32\linewidth, clip]{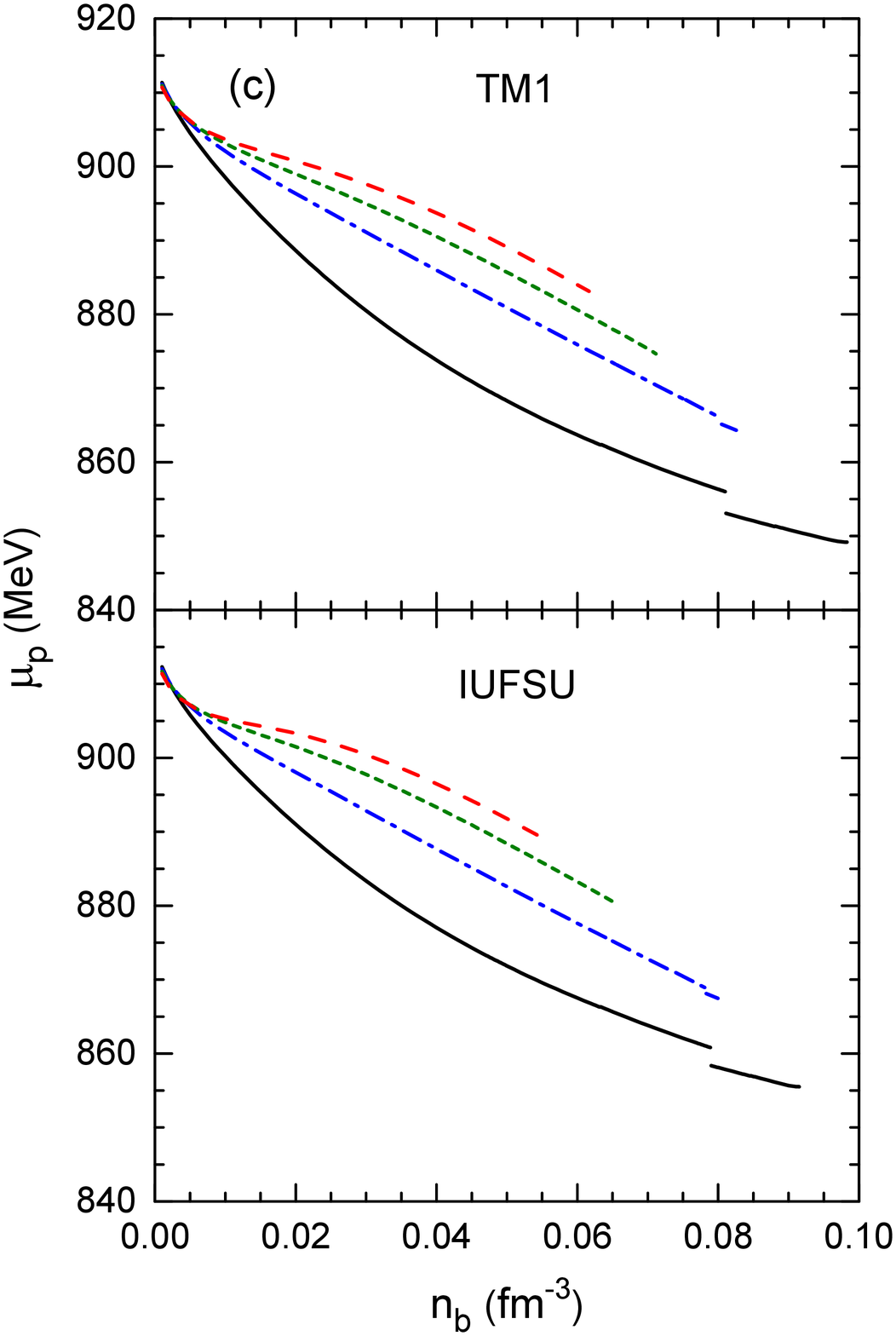}  \\
\end{tabular}
\caption{(Color online) Same as Fig.~\ref{fig:6TF}, but for chemical potentials
of electrons $\mu_e$ (a), neutrons $\mu_n$ (b), and protons $\mu_p$ (c).}
\label{fig:7mu}
\end{figure}
\end{center}

To examine the influence of $L$ on properties of the inner crust,
we perform self-consistent TF calculations using the two sets of
generated models based on the TM1 and IUFSU parametrizations.
In Fig.~\ref{fig:6TF}, we present the following quantities:
(a) the average proton fraction $Y_p$;
(b) neutron densities of the liquid phase and the gas phase,
    $n_{n,L}$ and $n_{n,G}$, at the center or boundary of the cell; and
(c) the proton density of the liquid phase, $n_{p,L}$, at the center or boundary of the cell.
As one can see from Fig.~\ref{fig:6TF}(a), $Y_p$ decreases rapidly
with increasing $n_b$ at lower densities and shows a significant
$L$ dependence at higher densities.
It is found that a smaller $L$ corresponds to a larger $Y_p$ at
a fixed $n_b$. This trend is related to the density dependence of
$E_{\rm{sym}}$ shown in Fig.~\ref{fig:1esym}.
Since $E_{\rm sym}$ has been fixed at $n_b$ = 0.11 fm$^{-3}$,
a smaller $L$ in one set of generated models corresponds to a larger
$E_{\rm sym}$ at lower densities ($n_b<$ 0.11 fm$^{-3}$).
It is well known that a larger $E_{\rm sym}$ favors a higher $Y_p$
in homogeneous $\beta$-equilibrium matter.
Therefore, a smaller $L$ results in a larger $Y_p$ at the density
close to the transition to homogeneous matter.
The correlation between $L$ and $Y_p$ is consistent with
those reported in Refs.~\cite{Oyam07,Gril12}.
A clear $L$ dependence is also observed in Fig.~\ref{fig:6TF}(b),
in which a smaller $L$ corresponds to larger $n_{n,L}$ and smaller
$n_{n,G}$. The $L$ dependence of $n_{n,L}$ and $n_{n,G}$
can be explained by the density dependence of $E_{\rm sym}$
as discussed above. It is shown in Fig.~\ref{fig:1esym} that
a smaller $L$ corresponds to larger $E_{\rm sym}$
at $n_b<$ 0.11 fm$^{-3}$ and smaller $E_{\rm sym}$
at $n_b>$ 0.11 fm$^{-3}$. Therefore, the model with a smaller $L$
favors more neutrons in the liquid phase and fewer neutrons in the gas phase,
which results in larger $n_{n,L}$ and smaller $n_{n,G}$, as shown in
Fig.~\ref{fig:6TF}(b). We note that the behaviors of $n_{n,L}$ and
$n_{n,G}$ obtained in the present study are consistent with those
reported in Refs.~\cite{Oyam07,Gril12}.
On the other hand, the behavior of $n_{p,L}$
is somewhat complicated, as shown in Fig.~\ref{fig:6TF}(c).
The model with a smaller $L$ results in a more rapid decrease of
$n_{p,L}$ at 0.01 $<n_b<$ 0.05 fm$^{-3}$.
This trend may be related to
the $\omega$-$\rho$ coupling term and behaviors of chemical
potentials as discussed in Ref.~\cite{Shen14}.
In Fig.~\ref{fig:7mu}, we display chemical potentials of
electrons, $\mu_e$, neutrons, $\mu_n$, and protons, $\mu_p$,
as a function of $n_b$ obtained in the TF approximation
for the two sets of generated models with several values of $L$.
The chemical potentials are intensive quantities that play
an important role in the TF calculation.
With increasing $n_b$, both $\mu_e$ and $\mu_n$ increase
monotonically in all cases of $L$, whereas $\mu_p$ decreases.
For the $L$ dependence of chemical
potentials at a fixed $n_b$, it is found that the model with
a smaller $L$ generally results in larger $\mu_e$ and $\mu_n$,
as well as smaller $\mu_p$. This can be explained by the
contribution from the $\rho$ meson in the RMF model. According to
Eqs.~(\ref{eq:mup}) and (\ref{eq:mun}), the value of $g_\rho \rho$
plays an essential role in determining the chemical potentials
$\mu_p$, $\mu_n$, and $\mu_e=\mu_n-\mu_p$.
One can see from Tables~\ref{tab:3} and~\ref{tab:4} that the model
with a smaller $L$ has relatively larger $g_{\rho}$ and $\Lambda_{\rm{v}}$,
which yields a larger value of $g_\rho \rho$.
Therefore, the model with a smaller $L$ leads to larger $\mu_e$
and $\mu_n$, as well as smaller $\mu_p$.
By comparing results between TM1 and IUFSU, we find that
the two sets of generated models have similar $L$ dependence for
all properties mentioned above. These results are also consistent
with those reported in Refs.~\cite{Gril12,Bao14,Oyam07,Shen14}.

\begin{figure}[htb]
\includegraphics[bb=55 36 420 829, width=7 cm,clip]{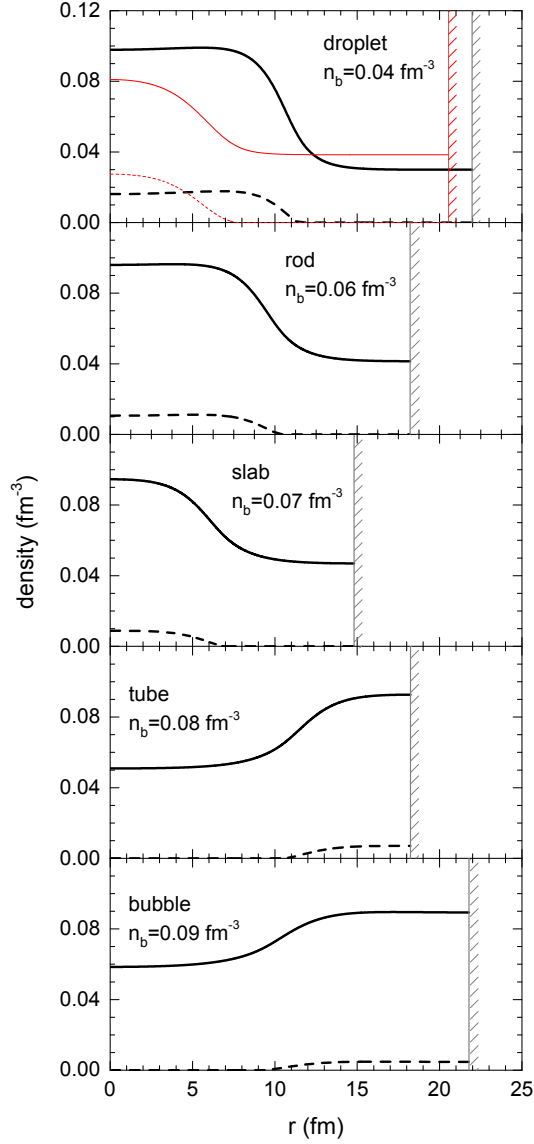}
\caption{(Color online) Density distributions of neutrons (solid lines) and protons
(dashed lines) in the Winger--Seitz cell at $n_b=0.04,\,0.06,\,0.07,\,0.08,$
and 0.09 fm$^{-3}$ (top to bottom) obtained in the TF approximation.
The results with the original IUFSU model ($L=47.2$ MeV) are shown
by thick black lines. For comparison, the results with the largest $L$
in the set of IUFSU ($L=110$ MeV) are shown by thin red lines in the top panel.
The cell boundary is indicated by the hatching.}
\label{fig:8dis}
\end{figure}

It is interesting to see how the density profile evolves through
various pasta phases as the density increases.
In Fig.~\ref{fig:8dis}, the density distributions of neutrons and
protons inside the Wigner--Seitz cell are plotted at several values
of $n_b$.
The calculations are performed with two extreme values
of $L$ in the set of IUFSU ($L=47.2$ MeV and $L=110$ MeV).
As shown in Fig.~\ref{fig:2diag}, the original IUFSU model ($L=47.2$ MeV)
predicts that all pasta phases would occur before the crust-core
transition, whereas only the droplet configuration appears in the
case of $L=110$ MeV. In Fig.~\ref{fig:8dis}, from top to bottom,
we show the results of droplet, rod, slab, tube, and bubble,
respectively. In the top panel at $n_b$ = 0.04 fm$^{-3}$,
the results for a droplet with $L=47.2$ MeV (thick lines) are
compared to those with $L=110$ MeV (thin lines).
It is seen that the droplet size obtained with $L=47.2$ MeV is
larger than that with $L=110$ MeV, and the nucleon
distributions in the case of $L=110$ MeV are more diffuse than those
of $L=47.2$ MeV. Moreover, as compared to the case of $L=110$ MeV,
the original IUFSU model with $L=47.2$ MeV yields lower neutron gas
density at the boundary ($n_{n,G}$) and higher neutron density at the
center of the cell ($n_{n,L}$), which are consistent with those shown in
Fig.~\ref{fig:6TF}(b). In the second panel at $n_b$ = 0.06 fm$^{-3}$,
the original IUFSU model predicts a rod phase,
while the model with $L=110$ MeV has undergone the crust-core transition.
The transition from slab to tube occurs at $n_b$ $\sim$ 0.079 fm$^{-3}$
in the original IUFSU model, and, therefore, sudden changes in the density
distributions are observed by comparing the third and fourth panels;
these cause visible jumps in chemical potentials and $Y_p$ as shown above.
As one can see, from top to bottom, the distributions of neutrons and protons
become more diffuse with increasing $n_b$, and the neutron density
difference between the liquid phase and the gas phase becomes
smaller and smaller. The crust-core transition occurs
at $n_b$ $\sim$ 0.092 fm$^{-3}$ in the original IUFSU model.

Finally, we discuss the correlation between the symmetry energy slope $L$
and the crust-core transition. In Fig.~\ref{fig:9PT}, we display the
crust-core transition density $n_{b,t}$ and the proton fraction and
the pressure at the transition point, $Y_{p,t}$ and $P_t$, as a function
of $L$ obtained from the self-consistent TF calculation using the two
sets of generated models based on the TM1 and IUFSU parametrizations.
As one can see from Fig.~\ref{fig:9PT}(a),
there is a clear correlation between $L$ and $n_{b,t}$; namely,
$n_{b,t}$ decreases monotonically with increasing $L$.
This correlation is consistent with those reported
in Refs.~\cite{Gril12,Oyam07,Bao14,Duco11}. Compared to the results
obtained using the CP method~\cite{Bao14}, the transition densities
obtained from the present TF calculation are slightly higher,
which may be related to the differences in configuration space
and treatments of surface and Coulomb energies.
From Fig.~\ref{fig:9PT}(b), it is seen that $Y_{p,t}$ decreases
significantly with increasing $L$. This trend is very similar to
that observed in Refs.~\cite{Bao14,Duco11}.
It can be understood from the density dependence of $E_{\rm{sym}}$
shown in Fig.~\ref{fig:1esym}. As mentioned above, a smaller $L$ in one
set of generated models corresponds to a larger $E_{\rm sym}$ at
$n_b<$ 0.11 fm$^{-3}$, and a larger $E_{\rm sym}$ favors a
higher $Y_p$ in homogeneous $\beta$-equilibrium matter.
Therefore, a smaller $L$ results in a larger $Y_{p,t}$ as
illustrated in Fig.~\ref{fig:9PT}(b).
The $L$ dependence of the transition pressure $P_t$ is nonmonotonic,
as shown in Fig.~\ref{fig:9PT}(c). It is found that $P_t$ decreases
with increasing $L$ in the large-$L$ region ($L>60$ MeV), whereas the
opposite behavior is observed for $L<60$ MeV.
A similar behavior was also observed in Ref.~\cite{Bao14}.
The nontrivial $L$ dependence of $P_t$ is due to
a competing effect, as discussed in Refs.~\cite{Bao14,Duco10}.
For neutron-rich matter at fixed density and proton fraction,
the pressure would increase with increasing $L$. However,
the decrease of $n_{b,t}$ shown in Fig.~\ref{fig:9PT}(a)
leads to a decrease of the pressure with increasing $L$.
As a result, $P_t$ depends on $L$ nonmonotonically, as
illustrated in Fig.~\ref{fig:9PT}(c).

\section{Conclusions}
\label{sec:4}

In this study, we have investigated the impact of the symmetry energy
on pasta phase properties within the self-consistent TF approximation.
It has been found that the symmetry energy slope $L$
plays an important role in determining the pasta phase structure and
the crust-core transition. In order to clarify the influence of $L$,
we have employed two sets of generated models based on the TM1 and IUFSU
parametrizations that have the same isoscalar saturation properties
and fixed symmetry energy at the density $n_b$ = 0.11 fm$^{-3}$
but have different symmetry energy slope $L$.
It has been observed that the model with a smaller $L$ predicts
an earlier onset of nonspherical nuclei and a later transition
to homogeneous matter. For example, the original IUFSU model with
$L=47.2$ MeV predicts that the transition from droplet to rod occurs
at $n_b$ $\sim$ 0.049 fm$^{-3}$, then the pasta phases of slab,
tube, and bubble appear one by one, and finally the transition
to homogeneous matter occurs at $n_b$ $\sim$ 0.092 fm$^{-3}$.
In contrast, the original TM1 model with $L=110.8$ MeV predicts
that only the droplet configuration appears at low density,
and the transition from droplet to homogeneous matter occurs
at $n_b$ $\sim$ 0.062 fm$^{-3}$. Therefore, it can be concluded
that the model with a smaller $L$ results in a more complex
phase structure and a larger density range of pasta phases
for neutron star crusts. In addition, some properties of the inner
crust such as the proton fraction, chemical potentials, and
density profiles have also been found to be correlated to the
symmetry energy slope $L$, which could be partly explained by
the density dependence of the symmetry energy.
We have compared the results of the present TF calculation
with those obtained using the CP method. It has been found that,
although there are quantitative differences between these two
methods, the qualitative behaviors of the inner crust are
very similar to each other, and they are consistent with
those reported in earlier studies~\cite{Oyam07,Gril12}.

The correlation between the symmetry energy slope $L$ and the
crust-core transition has been examined in the TF approximation
using the two sets of models generated from the TM1 and IUFSU
parametrizations. It has been found that the crust-core transition
density $n_{b,t}$ decreases monotonically with increasing $L$.
This correlation is consistent with those reported
in Refs.~\cite{Gril12,Oyam07,Bao14,Duco11}.
Also, the proton fraction at the transition point $Y_{p,t}$
decreases significantly with increasing $L$, whereas the
transition pressure $P_t$ shows a nontrivial dependence on $L$.
These correlations obtained from the present TF calculation are
consistent with those obtained using the CP method~\cite{Bao14}.
In the TF approximation, nuclear shell and pairing effects
have been neglected. It is of interest to further consider neutron
paring and superfluidity in the inner crust of neutron stars.


\section*{Acknowledgment}

This work was supported in part by the National Natural Science Foundation
of China (Grant No. 11375089).


\newpage
\begin{table}[tbp]
\caption{Parameter sets used in this work. The masses are given in MeV.}
\begin{center}
\begin{tabular}{lccccccccccc}
\hline\hline
Model   &$M$  &$m_{\sigma}$  &$m_\omega$  &$m_\rho$  &$g_\sigma$  &$g_\omega$
        &$g_\rho$ &$g_{2}$ (fm$^{-1}$) &$g_{3}$ &$c_{3}$ &$\Lambda_{\textrm{v}}$ \\
\hline
TM1     &938.0  &511.198  &783.0  &770.0  &10.0289  &12.6139  &9.2644
        &$-$7.2325   &0.6183   &71.3075   &0.000  \\
IUFSU   &939.0  &491.500  &782.5  &763.0  &9.9713   &13.0321  &13.5900
        &$-$8.4929   &0.4877   &144.2195  &0.046 \\
\hline\hline
\end{tabular}
\label{tab:1}
\end{center}
\end{table}

\begin{table}[htb]
\caption{Saturation properties of nuclear matter for the TM1 and IUFSU models.
The quantities $E_0$, $K$, $E_{\text{sym}}$, and $L$ are, respectively,
the energy per nucleon, incompressibility coefficient, symmetry
energy, and symmetry energy slope at saturation density $n_0$.}
\label{tab:2}
\begin{center}
\begin{tabular}{l c c c c c}
\hline\hline
Model & $n_0$ (fm$^{-3}$) & $E_0$ (MeV) & $K$ (MeV) & $E_{\text{sym}}$ (MeV) & $L$ (MeV) \\
\hline
TM1   & 0.145 & $-$16.3 & 281.0 & 36.9 & 110.8 \\
IUFSU & 0.155 & $-$16.4 & 231.0 & 31.3 & 47.2  \\
\hline\hline
\end{tabular}
\end{center}
\end{table}

\begin{table}[htb]
\caption{Parameters, $g_\rho$ and $\Lambda_{\text{v}}$, generated from the TM1 model
for different slope $L$ at saturation density $n_0$ with fixed symmetry energy
$E_{\text{sym}}=28.05$ MeV at a density of 0.11 fm$^{-3}$.
The last line shows the symmetry energy at saturation density,
$E_{\text{sym}} (n_0)$. The original TM1 model has $L=110.8$ MeV.}
\label{tab:3}
\begin{center}
\begin{tabular}{lcccccccc}
\hline\hline
$L$ (MeV) & 40.0    & 50.0    & 60.0    & 70.0    & 80.0    & 90.0   & 100.0  & 110.8  \\
\hline
$g_\rho$  & 13.9714 & 12.2413 & 11.2610 & 10.6142 & 10.1484 & 9.7933 & 9.5114 & 9.2644 \\
$\Lambda_{\text{v}}$ & 0.0429 & 0.0327  & 0.0248  & 0.0182  & 0.0128 & 0.0080 & 0.0039 & 0.0000 \\
$E_{\text{sym}} (n_0)$ (MeV) & 31.38 & 32.39  & 33.29   & 34.11   & 34.86  & 35.56  & 36.22  & 36.89 \\
\hline\hline
\end{tabular}%
\end{center}
\end{table}

\begin{table}[htb]
\caption{Parameters, $g_\rho$ and $\Lambda_{\text{v}}$, generated from the IUFSU model
for different slope $L$ at saturation density $n_0$ with fixed symmetry energy
$E_{\text{sym}}=26.78$ MeV at a density of 0.11 fm$^{-3}$.
The last line shows the symmetry energy at saturation density,
$E_{\text{sym}} (n_0)$. The original IUFSU model has $L=47.2$ MeV.}
\label{tab:4}
\begin{center}
\begin{tabular}{lcccccccc}
\hline\hline
$L$ (MeV) & 47.2    & 50.0    & 60.0    & 70.0    & 80.0   & 90.0   & 100.0  & 110.0  \\
\hline
$g_\rho$  & 13.5900 & 12.8202 & 11.1893 & 10.3150 & 9.7537 & 9.3559 & 9.0558 & 8.8192 \\
$\Lambda_{\text{v}}$& 0.0460  & 0.0420  & 0.0305  & 0.0220 & 0.0153 & 0.0098 & 0.0051 & 0.0011 \\
$E_{\text{sym}} (n_0)$ (MeV) & 31.30 & 31.68  & 32.89   & 33.94  & 34.88  & 35.74  & 36.53  & 37.27  \\
\hline\hline
\end{tabular}%
\end{center}
\end{table}

\begin{table}[htb]
\caption{Comparison of the onset densities of pasta phases and homogeneous
matter (hom.) between the TF and CP methods for the original IUFSU
and TM1 models.}
\begin{center}
\begin{tabular}{lccccccc}
\hline\hline
Model & Method & \multicolumn{6}{c}{Onset density (fm$^{-3}$)}      \\
\cline{3-8}
      &    &droplet &rod    &slab   &tube   &bubble & hom.    \\
\hline
IUFSU & TF &0.000  &0.049 &0.063 &0.079 &0.085 &0.092  \\
IUFSU & CP &0.003  &0.055 &0.070 &0.088 & ---  &0.089  \\
TM1   & TF &0.000  & ---  & ---  & ---  & ---  &0.062  \\
TM1   & CP &0.001  & ---  & ---  & ---  & ---  &0.058  \\
\hline\hline
\end{tabular}
\label{tab:5}
\end{center}
\end{table}

\newpage

\begin{figure}[htb]
\includegraphics[bb=11 10 547 757, width=7 cm,clip]{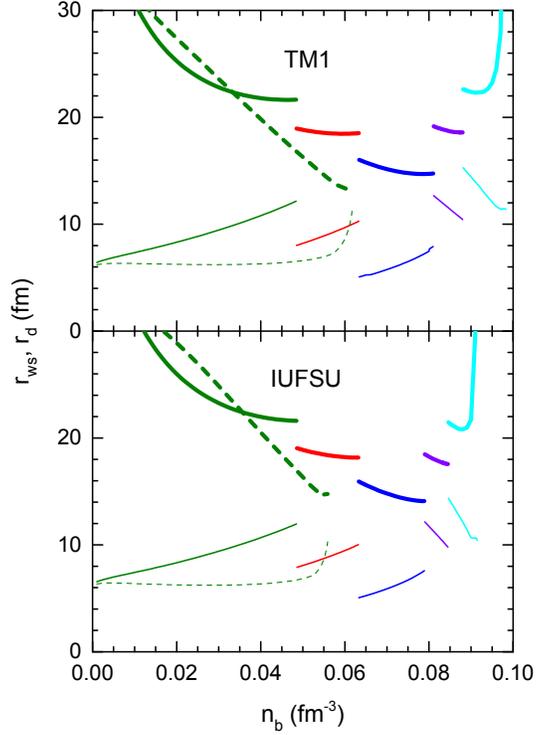}
\caption{(Color online) Size of the Wigner--Seitz cell, $r_{\text{ws}}$
(thick lines), and that of the inner part, $r_d$ (thin lines),
as a function of $n_b$ obtained in the TF approximation.
The results calculated with the smallest $L$ in TM1
($L=40$ MeV) and IUFSU ($L=47.2$ MeV) are shown by solid lines,
where the pasta phase structure changes from droplet (green) to
rod (red), slab (blue), tube (violet), and bubble (cyan)
as the density increases. For comparison, the results
with the largest $L$ in TM1 ($L=110.8$ MeV)
and IUFSU ($L=110$ MeV) are shown by green-dashed lines,
where only the droplet configuration appears before the crust-core
transition.}
\label{fig:4rcd}
\end{figure}

\end{document}